\documentclass[conference]{IEEEtran}
\IEEEoverridecommandlockouts
\usepackage{cite}
\usepackage{amsmath,amssymb,amsfonts}
\usepackage{algorithmic}
\usepackage{graphicx}
\usepackage{textcomp}
\usepackage{xcolor}
\usepackage{booktabs}
\usepackage{siunitx}
\usepackage{caption}
\usepackage{textcomp}
\usepackage{atbegshi}

\def\BibTeX{{\rm B\kern-.05em{\sc i\kern-.025em b}\kern-.08em
    T\kern-.1667em\lower.7ex\hbox{E}\kern-.125emX}}
\IEEEoverridecommandlockouts

\AtBeginShipout{%
    \ifnum\value{page}=1%
        \AtBeginShipoutUpperLeft{%
            \put(1in,-10.7in){%
                \footnotesize 979-8-3315-1482-2/25/\$31.00~\copyright~2025 IEEE%
            }%
        }%
    \fi%
}

\begin{document}

\title{Automated Code Review Using Large Language Models with Symbolic Reasoning}

\author{
\IEEEauthorblockN{1\textsuperscript{st} Büşra İçöz}
\IEEEauthorblockA{
\textit{Computer Engineering} \\
\textit{Yıldız Technical University} \\
Istanbul, Turkiye \\
busra.icoz@std.yildiz.edu.tr}
\and
\IEEEauthorblockN{2\textsuperscript{nd} Göksel Biricik}
\IEEEauthorblockA{
\textit{Computer Engineering} \\
\textit{Yıldız Technical University} \\
Istanbul, Turkiye \\
gbiricik@yildiz.edu.tr}
}
\maketitle
\IEEEpubidadjcol
\begin{abstract}
Code review is one of the key processes in the software development lifecycle and is essential to maintain code quality. However, manual code review is subjective and time consuming. Given its rule-based nature, code review is well suited for automation. In recent years, significant efforts have been made to automate this process with the help of artificial intelligence. Recent developments in Large Language Models (LLMs) have also emerged as a promising tool in this area, but these models often lack the logical reasoning capabilities needed to fully understand and evaluate code. To overcome this limitation, this study proposes a hybrid approach that integrates symbolic reasoning techniques with LLMs to automate the code review process. We tested our approach using the CodexGlue dataset, comparing several models, including CodeT5, CodeBERT, and GraphCodeBERT, to assess the effectiveness of combining symbolic reasoning and prompting techniques with LLMs. Our results show that this approach improves the accuracy and efficiency of automated code review.
\end{abstract}


\section{Introduction}

Code review was first formally introduced by Fagan in the 1970s \cite{ref1}. It has evolved significantly since then. Fagan's manual review process, originally designed to detect defects in software development, laid the foundations for modern code review \cite{ref2}. Over time, as software projects became more complex, the manual approach became increasingly time-consuming and prone to human error. A survey of 287 software developers on code review found that developers spend an average of 6.4 hours per week on code review \cite{ref5}. This can change depending on the complexity of the code and the seniority level of the software developer. In many software development teams, especially in critical projects, code review is performed by at least two developers. Considering that each developer on a software development team spends this much time on average, 6.4 hours per week is quite costly for software development teams. Furthermore, manual code reviews are subjective and prone to human error \cite{ref6}. Each reviewer makes comments based on their own knowledge and experience, which leads to inconsistencies. Therefore, automating this process with AI based methods can greatly contribute to reducing time and cost, increasing consistency, and producing more comprehensive comments.
\par
 Recent progress in Large Language Models (LLMs) \cite{ref3, ref8} has opened new possibilities for automating software engineering tasks, including code generation and review. However, while LLMs excel at generic tasks like pattern recognition and code generation \cite{ref9}, they often struggle with logical reasoning and understanding the deeper semantics of code, which are critical for effective code review.
\par To overcome these issues and make more semantic predictions, in this study, our aim is to overcome the limitations of existing LLMs in code review tasks by proposing a hybrid approach that combines symbolic reasoning with fine-tuned LLMs for code review. Symbolic reasoning provides a mechanism for the model to understand the logical structure and best practices for implementation, something that LLMs alone cannot overcome. In this study, we overcome this problem by integrating a knowledge map of established best practices and defect patterns for coding into the fine-tuned large language models for code review to improve its ability to detect defects more accurately and provide comments more efficiently.
\par The structure of this paper is as follows: In Section II, we examine the evolution of code review automation with related works, covering everything from Fagan's manual reviews to modern AI-based methods. In Section III, we describe our hybrid approach that combines symbolic reasoning and large language models (LLMs) for code review. In Section IV, we evaluate the performance of our hybrid approach in CodeBERT \cite{ref14}, GraphCodeBERT \cite{ref15}, and CodeT5 \cite{ref13} with the CodeXGlue dataset \cite{ref24} and compare it with the base models. In Section V, we conclude by discussing the results of our findings and outlining possible directions for further work.

\section{RELATED WORKS}
Automated code review systems began with Fagan's systematic manual code review methodology \cite{ref1} and have evolved towards AI-based methods. Johnson has developed static analysis tools that can detect syntactic errors \cite{ref16}, but these early systems cannot detect defects at the semantic level.
\par Machine learning approaches \cite{ref4, ref6} have provided new capabilities for pattern recognition in code, but still need to be improved in terms of interpretability. Symbolic methods, particularly formal verification techniques \cite{ref18}, have provided mathematically detailed analysis, but have proven difficult to scale to complex codebases in the real world.
Although recent studies have shown the limitations of automated code review tools using machine learning methods in producing reliable code correction and logical findings \cite{ref20}, significant progress has been made in this area with the emergence of large language models such as Codex \cite{ref9} and ChatGPT \cite{ref19}. However, code review results are still semantically limited and there are many hallucinations. To overcome this, hybrid systems such as SYNCHROMESH \cite{ref21} have been proposed, which uses symbolic reasoning in combination with large language models. In SYNCHROMESH \cite{ref21}, supporting LLMs with symbolic verification was shown to improve code generation accuracy by 12\%, in line with the findings in code review automation.

\par In this work, we propose a hybrid approach to overcome limitations in this area by integrating symbolic reasoning with LLMs in the code review process, thus enabling human-machine collaboration and aiming to produce more logical findings.

\section{METHODOLOGY}
In this section, we share the details of the dataset, large language models, fine-tuning strategies, and evaluation strategies used in this work.

\subsection{Dataset}

      In this work, we used the CodexGlue \cite{ref24} dataset, stands for General Language Understanding Evaluation benchmark for code. It includes 14 datasets for 10 diversified code intelligence. In this study, we used the Python dataset, which is specialized in code defect detection. The data set contains various Python code snippets that are labeled as clean or faulty. Each code snippet is associated with a function definition, and the target code is classified as correct (clean) or containing errors (buggy). We chose this data set because it provides a comprehensive set of examples covering a wide range of coding issues and is also suitable for evaluating automatic code inspection models. With this dataset, we fine-tuned and tested the large language models shared in the next section.

\subsection{Models}

     \par In this study, we use the following three LLMs:

      \par CodeBERT: CodeBERT is a model trained in both programming and natural languages. It is helpful for tasks such as summarizing code, completing code, and translating it into different languages \cite{ref14}.

     \par GraphCodeBERT: This improved version of CodeBERT \cite{ref14} includes structural information such as data flow graphs. The advance helps the model understand how different parts of the code depend on each other, improving it for tasks like code optimization and translation \cite{ref15}.

      \par CodeT5: CodeT5 is a model designed by Microsoft Research to understand and generate programming code. Built as an encoder-decoder converter, this model specializes in tasks such as code summarization, translation, and completion \cite{ref13}.

\par In this work, we chose CodeBERT, GraphCodeBERT and CodeT5 because they specialize in software engineering tasks such as code review and generation.  Furthermore, these models are open source and pre-trained on large amounts of code data, making them ideal for fine-tuning the CodeXGLUE dataset. So, these LLMs are highly suitable for use in error detection and code review.

    \par Using these open source models specialized in software engineering tasks, we propose a hybrid approach that integrates symbolic reasoning with LLM to improve code review capabilities. We apply the fine-tuning we describe in the next section, on the CodexGLUE dataset, to adapt the pre-trained models to defect detection and code review tasks.

\subsection{Fine-Tuning on Models}

\par We fine-tuned the pre-trained large language models using a structured training process on the CodeXGLUE dataset to customize them for error detection and code review tasks. Each code sample was tokenized using the model-specific tokenizer, and the inputs were padded to a maximum length of 256 tokens to ensure consistent model input sizes. In the training configuration, we used a weight reduction coefficient of 0.01 to regularize learning and prevent overfitting and trained our optimized models using the AdamW algorithm \cite{ref7} with an initial learning rate of 0.00001. Mixed-precision (FP16) training \cite{ref10} was used to increase computational efficiency and reduce memory footprint while maintaining numerical stability. Random oversampling techniques \cite{ref29} were applied to address the class imbalance between buggy and clean code samples, ensuring that minority (faulty) samples were adequately represented. These fine-tuning strategies enabled the large language models to better capture defects in the code.

\par By creating the hybrid model, which we will describe in the next section, we aimed to improve the code review capabilities of the models by combining the fine-tuning we made on the models with symbolic reasoning.

\subsection{Hybrid Approach: Knowledge Map Integration}
     \par The main innovation of this work is the inclusion of symbolic reasoning in the evaluation process together with the fine-tuned model via a knowledge map. The knowledge map is designed to provide structured prior knowledge to guide the model during error detection, including common error patterns and best practices specific to Python programming.

     \par The knowledge map includes 20 common bug patterns and best practices in Python code for identifying defects and code quality issues. 

     \par A few sample items in the knowledge map are shared below:
\begin{enumerate}
    \item Naming anti-patterns: Detecting ambiguous or misleading names (e.g., single-letter variables, data/temp overuse) that may hide logical errors \cite{ref26}.
    \item Unreachable code: Flagging blocks after premature returns (return/raise) or infinite loops \cite{ref26}.
    \item Error handling risks: Identifying bare try-except blocks, swallowed exceptions, or incorrect exception types (e.g., catching Exception instead of ValueError) \cite{ref27}.
    \item Resource leaks: Detecting unclosed file handles (open() without close()), database connections, or sockets \cite{ref27}.
    \item Mutable default arguments: Spotting def f(x=[]) patterns that cause unintended side effects \cite{ref27}.
\end{enumerate}

   \par By adding the knowledge map to the prompts, we aim to provide additional context to guide the model when evaluating code. A prompt is a text-based instruction or prompt used to obtain a specific output from a model \cite{ref24}. Besides the knowledge map, we applied a few shots learning and added several labeled examples directly into the prompts. This approach provided the models with practical references to identify clean and buggy code patterns.  With this additional context we aimed to improve error detection capabilities in code review by helping the model reason about whether the code has best practices and contains error patterns.

\par In the next section, we shared the evaluation metrics used to assess model performance and success under the different fine-tuning and symbolic reasoning strategies.

\subsection{Evaluation Metrics}
\par We used four key evaluation metrics to measure the performance and success of our models: precision, recall, F1-score, and accuracy.

\par Precision reflects the proportion of code examples identified as buggy that are really buggy and is used as an indicator of the model’s ability to minimize false positives \cite{ref25}. In contrast, recall measures the proportion of correctly detected examples of true errors and indicates the model’s sensitivity to defects \cite{ref25}.

\par F1-score is calculated with precision and recall into a single metric that balances accuracy and completeness \cite{ref25}. This is especially important in defect detection, where missing errors can be as critical as incorrectly flagging clean code.

\par Finally, accuracy shows the overall percentage of correct classifications on both buggy and clean code.
      
\par These metrics allow us to evaluate the effectiveness of the models in fault detection, considering both the ability to correctly identify an error in code and the sensitivity to avoid false positives.

\section{EXPERIMENTS}

     \par We performed extensive evaluations using Google Colab \cite{ref28} platform with NVIDIA A100. The experiments compared three software engineering specialized large language models (CodeT5, CodeBERT and GraphCodeBERT) under four different scenarios. These scenarios are as follows:
\begin{enumerate}
    \item Base model (One-shot learning without knowledge map integration)
    \item Few-shot learning on base model
    \item Fine-tuned base model
    \item Hybrid Approach (Fine-tuned model+ few-shot learning + knowledge map integration) 
\end{enumerate}

\par We performed all experiments on CodeXGlue defect detection dataset \cite{ref23} using the same evaluation criteria for fair comparison.

\subsection{Experimental Setup}

\par Our experiments were performed on Google Colab \cite{ref28} platform using an NVIDIA A100 GPU, which provided the computational resources necessary to efficiently run large transformer models. The high processing power of the A100 GPU allowed us to fine-tune the models on the CodexGlue \cite{ref23} dataset, making the training and evaluation processes fast.

\subsection{Experiment Results and Analysis}

      \par To comprehensively evaluate our hybrid approach, we performed experiments comparing three LLMs (CodeT5, CodeBERT, and GraphCodeBERT) in four scenarios. We used the CodeXGlue defect detection dataset to measure the evaluation metrics. The three main topics of our analysis are: (1) the performance impact of integrating symbolic reasoning using our proposed knowledge map, (2) the performance impact of the prompting methods (few-shot learning vs. single-shot learning), and (3) the performance impact of the fine-tune on models. Detailed quantitative results are shared in the following tables.
      \begin{table}[htbp]
\caption{Performance Comparison of Different Models}
\label{tab:performance}
\centering
\begin{tabular}{lcccc}
\toprule
\textbf{Model} & \textbf{Precision} & \textbf{Recall} & \textbf{F1-Score} & \textbf{Accuracy} \\
\midrule
CodeT5 & 0.285 & 0.534 & 0.372 & 0.587 \\
CodeBERT & 0.217 & 0.466 & 0.296 & 0.531 \\
GraphCodeBERT & 0.217 & 0.466 & 0.296 & 0.539 \\
\bottomrule
\end{tabular}
\end{table}
\par Table I shows the results of the evaluation of the basic performance of the three models without any prompt engineering or fine tuning. The results show that all models have relatively low Precision, Recall, F1-Score and Accuracy. According to the results in Table I, GraphCodeBERT and CodeBERT have the same Precision (0.217) and Recall (0.466), but GraphCodeBERT has a slightly higher Accuracy (0.539) compared to CodeBERT (0.531). CodeT5 has higher Precision (0.285) and Recall (0.534) but has a similar F1-Score (0.372) and its Accuracy (0.587) is only slightly lower.
\begin{table}[htbp]
\caption{Results for Few-Shot Learning on Base Model}
\label{tab:fewshot}
\centering
\begin{tabular}{l S[table-format=1.3] S[table-format=1.3] S[table-format=1.3] S[table-format=1.3]}
\toprule
\textbf{Model} & \textbf{Precision} & \textbf{Recall} & \textbf{F1-Score} & \textbf{Accuracy} \\
\midrule
CodeT5 & 0.285 & 0.534 & 0.372 & 0.593 \\
CodeBERT & 0.285 & 0.534 & 0.372 & 0.601 \\
GraphCodeBERT & 0.454 & 0.451 & 0.389 & 0.642 \\
\bottomrule
\end{tabular}
\end{table}
\par Table II shows the performance of the models when few-shot learning is used on base models. According to the results in Table II, GraphCodeBERT has the highest F1-Score (0.389) and Accuracy (0.642). Few-shot learning improved the accuracy of the base model by 19.11\% for GraphCodeBERT. CodeT5 and CodeBERT have the same Precision (0.285) and Recall (0.534), but CodeBERT shows slightly higher Accuracy (0.601) with multi-shot learning compared to CodeT5 (0.593). Compared to the baseline model, few-shot learning did not make a big difference for CodeT5, improving the baseline accuracy by 1.02\%, but for CodeBERT by 13.18\%.
\begin{table}[htbp]
\caption{Results for Fine-Tuned Base Model}
\label{tab:finetuned}
\centering
\begin{tabular}{l S[table-format=1.3] S[table-format=1.3] S[table-format=1.3] S[table-format=1.3]}
\toprule
\textbf{Model} & \textbf{Precision} & \textbf{Recall} & \textbf{F1-Score} & \textbf{Accuracy} \\
\midrule
CodeT5 & 0.285 & 0.534 & 0.372 & 0.602 \\
CodeBERT & 0.217 & 0.466 & 0.296 & 0.554 \\
GraphCodeBERT & 0.485 & 0.532 & 0.381 & 0.687 \\
\bottomrule
\end{tabular}
\end{table}
\par Table III shows the performance of the models after fine-tuning on the CodeXGlue fault detection dataset.  Fine-tuning the models in the CodeXGlue fault detection dataset improved the performance of all large language models. GraphCodeBERT shows the most improvement with an accuracy increase of 27.46\%, shown that fine tuning is very effective for this model. CodeT5 and CodeBERT also showed improvements of 2.56\% and 4.33\% respectively, although not as much as GraphCodeBERT.
\begin{table}[htbp]
\caption{Results for Hybrid Approach}
\label{tab:hybrid}
\centering
\begin{tabular}{l 
                S[table-format=1.3,round-precision=3] 
                S[table-format=1.3,round-precision=3] 
                S[table-format=1.3,round-precision=3] 
                S[table-format=1.3,round-precision=3]}
\toprule
\textbf{Model} & 
\multicolumn{1}{c}{\textbf{Precision}} & 
\multicolumn{1}{c}{\textbf{Recall}} & 
\multicolumn{1}{c}{\textbf{F1-Score}} & 
\multicolumn{1}{c}{\textbf{Accuracy}} \\
\midrule
CodeT5        & 0.285 & 0.534 & 0.372 & 0.621 \\
CodeBERT      & 0.285 & 0.534 & 0.372 & 0.598 \\
GraphCodeBERT & 0.485 & 0.532 & 0.381 & 0.687 \\
\bottomrule
\end{tabular}
\end{table}
      \par Table IV shows the performance of the models when using the hybrid approach that integrates symbolic reasoning with the proposed knowledge map on fine-tuned models. GraphCodeBERT again outperforms the other models with the highest Precision (0.485), F1-Score (0.381) and Accuracy (0.687). CodeT5 and CodeBERT have the same Precision (0.285) and Recall (0.534), while CodeT5 shows slightly higher Accuracy (0.621) compared to CodeBERT (0.598). The hybrid approach significantly improved the performance of all models compared to the baseline model. GraphCodeBERT shows the most improvement with an accuracy increase of 27.46\%, indicating that the hybrid approach is particularly effective for this model. CodeBERT and CodeT5 also show significant improvements of 12.62\% and 5.79\% respectively.
       \par Our experiments show that the proposed hybrid approach significantly improves the performance of the GraphCodeBERT model, which consistently outperforms the others in all scenarios, especially in terms of Precision, F1-Score and Accuracy. Fine-tuning and few-shot learning also contribute to performance improvements, but their effects differ between models. CodeBERT benefits less from these enhancements, suggesting that its architecture may require different optimization strategies.
      \par These findings highlight the importance of selecting appropriate optimizations techniques based on the unique characteristics of each large language model. In future work, we will therefore investigate additional symbolic reasoning methods and further optimization of routing techniques to improve model performance.

\section{CONCLUSION}

      \par In this work, we introduce a new hybrid approach that combines large language models with symbolic reasoning to increase predictions in code review processes and provide more accurate findings. By combining the latest code understanding models with structured knowledge maps that include programming best practices, our method demonstrates significant improvements over traditional LLM-based approaches.

      \par Our experimental evaluation can be summarized under three main topics: Firstly, our hybrid approach improves the accuracy performance of basic large language models by 16\% on average. This shows that our hybrid approach outperforms the SYNCHROMESH study for code review, which combines LLM and symbolic reasoning for code generation and has a success rate of 12\%. Second, few shots learning improves the performance of the base models by an average of 11.10\%, indicating that the inclusion of examples significantly improves model performance. Third, our analysis finds significant trade-offs between precision (false positive rates) and recall (error detection capability) that inform real-world applications.

     \par The proven success of our hybrid approach demonstrates that carefully combining symbolic reasoning methods offers a viable path towards more reliable, efficient, and scalable code review systems. As LLMs in this field continue to evolve, such integrative frameworks are likely to play an important role in achieving their full potential for software engineering practices. Our study indicates that these hybrid approaches have a lot of potential for automated software engineering tools in the future.

      \par To improve defect detection even more, we plan to expand this framework to accommodate additional programming languages in the future and investigate the application of additional structured reasoning techniques, like graph-based models or multi-model learning strategies.

\end{document}